\documentclass[prl,twocolumn,showpacs,preprintnumbers,amsmath,amssymb,tightenlines,epsfig, superscriptaddress]{revtex4}

\usepackage{graphicx}
\usepackage{graphicx}
\usepackage{dcolumn}
\usepackage{bm}
\usepackage{epsfig}
\usepackage{stmaryrd}
\usepackage{enumerate}
\usepackage{color}
\usepackage{ulem}  
 \newcommand{\kbar}{\hbar_{\textrm{eff}}} 
 \newcommand{\real}[1]{\mathfrak{Re}\{#1\}}
 \newcommand{\imag}[1]{\mathfrak{Im}\{#1\}}
 
\newcommand{\sub}[2]{{#1}_{\mbox{\!\! \scriptsize #2}}}

\def\beq{\begin{align}}
\def\eeq{\end{align}}

\def\CR{\nonumber\\[0.15cm]}
\newcommand{\rref}[1]{Ref.~\cite{#1}}
\newcommand{\fref}[1]{Fig.~\ref{#1}}
\newcommand{\frefp}[2]{Fig.~\ref{#1}~(#2)}
\newcommand{\bref}[1]{(\ref{#1})}
\newcommand{\eref}[1]{Eq.~(\ref{#1})}
\newcommand{\erefs}[2]{Eqs.~(\ref{#1},\ref{#2})}

\begin{document}

\title{Macroscopic Quantum Self-Trapping  in Dynamical Tunnelling}
\author{Sebastian~W\"uster}
\affiliation{The University of Queensland, School of Mathematics and Physics, Brisbane, Qld 4072, Australia}
\affiliation{Max Planck Institute for the Physics of Complex Systems, N\"{o}thnitzer Strasse 38, 01187 Dresden, Germany}
\author{Beata~J.~D\c{a}browska-W\"uster}
\affiliation{The University of Queensland, School of Mathematics and Physics, Brisbane, Qld 4072, Australia}
\author{Matthew~J.~Davis}
\affiliation{The University of Queensland, School of Mathematics and Physics, Brisbane, Qld 4072, Australia}
\email{sew654@pks.mpg.de}

\begin{abstract}
It is well-known that increasing the nonlinearity due to repulsive atomic interactions in a double-well Bose-Einstein condensate suppresses quantum tunnelling between the two sites. Here we find analogous behaviour in the dynamical tunnelling of a Bose-Einstein condensate between period-one resonances in a single driven potential well.
For small nonlinearities we find unhindered tunnelling between the resonances, but with an increasing period as compared to the non-interacting system.  For nonlinearities above a critical value we generally observe that the tunnelling shuts down. However, for certain regimes of modulation parameters we find that dynamical tunnelling re-emerges for large enough nonlinearities, an effect not present in spatial double-well tunnelling.  We develop a two-mode model in good agreement with full numerical simulations over a wide range of parameters, which allows the suppression of tunnelling to be attributed to macroscopic quantum self-trapping.
\end{abstract}


\maketitle


The transition from the classical to the quantum world is a subject of intense interest.  In particular, the topic of quantum chaos studies  systems {which} exhibit chaotic dynamics {in the classical limit of $\hbar \rightarrow 0$}~\cite{schlosshauer:book,book:reichl,schlosshauer:review,buchleitner:fareytree,garciamata:entanglescreen}. An important phenomenon in  driven one-dimensional {quantum} systems is  dynamical tunnelling, first identified by Heller and Davis~\cite{heller:dt}.  This is a classically forbidden process whereby particles trapped in a regular region of \emph{phase space} may quantum-mechanically tunnel to another.  The behaviour of such systems has provided important insights into the quantum-classical transition~\cite{tomsovic:chaosassist,plata:classicalquantum,ballentine:tunneling,perez:tunneling,uterman:onsetofchaos,mouchet:signatures,eltschka:resonances,martin:matthew:chip}: in particular, the period of the dynamical tunnelling is strongly affected by a number of subtle effects~\cite{uterman:onsetofchaos,mouchet:signatures,eltschka:resonances,martin:matthew:chip}. 
Dynamical tunnelling has mostly been studied in the single-particle regime~\cite{hensinger:pra,mouchet:signatures,mouchet:resonances,hug:milburn,milburn:secondorderres, osovski:fingerprints}, and has been demonstrated experimentally with ultra-cold atoms in modulated optical lattice potentials~\cite{exp:hensinger,exp:raizen}. Recently it has been shown in \cite{martin:matthew:chip} that atomic interactions in trapped Bose-Einstein condensates (BECs) can have a detectable effect for experimentally realistic parameters.
Here we investigate the effect of repulsive atomic interactions on the dynamical tunnelling of a trapped BEC through the variation of the nonlinearity, $U$.  \textit{A priori}, the effect of nonlinearity on a given dynamical system is not clear.  It has been shown to suppress transport in the kicked rotor and oscillator~\cite{rebuzzini:nonlinear,rebuzzini:nonlinear2, rebuzzini:nonlinear:acc, wimberger:kickedcond}, {and} Landau-Zener tunnelling in optical lattices~\cite{wimberger:wannier:stark}, but may enhance~\cite{flach:ratchet} or suppress~\cite{heimsoth:orbitaljosephson} transport in quantum ratchets.  

We find that dynamical tunnelling {also} occurs for the interacting system with $U>0$ up to a critical interaction strength $\sub{U}{\textrm{crit}}$. Beyond $\sub{U}{\textrm{crit}}$ we find that  dynamical tunnelling mostly ceases.  We connect the dynamical tunnelling suppression to the phenomenon of macroscopic quantum self-trapping (MQST) using a two-mode model based on Floquet tunnelling states. It allows us to predict the critical nonlinearity $\sub{U}{\textrm{crit}}$ from knowledge of the noninteracting system, and to understand the increase of the tunnelling period with $U$ that we find numerically.
While previous work reported detrimental effects of nonlinearities on dynamical tunnelling~\cite{rebuzzini:nonlinear:dt}, a connection with MQST was not made.  {Surprisingly, at higher nonlinearities with $U > \sub{U}{\textrm{crit}}$ we find some  parameter ranges where dynamical tunnelling reappears. This effect has no analogue in bosonic Josephson junctions, where MQST   has been extensively studied~\cite{smerzi:mqst,holthaus:NLfloq,joel:MQST} and demonstrated experimentally~\cite{oberthaler:exp1}.}

We begin by reviewing {the} dynamical tunnelling of ultra-cold atoms.
For classical atoms in {a} one-dimensional (1D) potential to {exhibit chaotic dynamics,} the potential must be {both driven and} anharmonic. The experiments {demonstrating dynamical tunnelling} used a modulated sinusoidal potential provided by an optical lattice~\cite{exp:hensinger,exp:raizen}.
Here we instead consider the  dimensionless classical Hamiltonian
\begin{align}
H&=\frac{p^{2}}{2} + V(x,t)\equiv\frac{p^{2}}{2} + \kappa  \big[1 + \epsilon \cos(t) \big] \sqrt{1+ x^{2}}, 
\label{hamiltonian}
\end{align}
where $\kappa$ is the potential strength, $\epsilon$ the amplitude of the modulation, and $x$ and $p$ are position and momentum co-ordinates respectively.  Potentials {$V(x,t)$} as in \eref{hamiltonian} can be realised on atom-chip traps in the radial direction~\cite{martin:matthew:chip}. {This} potential has the conceptual advantage of not being periodic in space; however, the physics we describe below will be generic for any one-dimensional potential where dynamical tunnelling is realised.

For our \emph{quantum} treatment of the system we consider a BEC subjected to this single-particle Hamiltonian.  We assume that mean-field theory is valid and the BEC is well described by the wave function $\Psi \equiv \Psi(x)$ that evolves according to the Gross-Pitaevskii equation (GPE)~\cite{stringari:review}: $i\kbar \frac{\partial}{\partial t} \Psi =\left[H + U |\Psi|^{2} \right]\Psi$, with $\int dx |\Psi|^2$=1. Here $p$ of \eref{hamiltonian} becomes $p=-i\kbar\frac{\partial}{\partial x}$ and consequently $[x,p]=i\kbar$.
 $U$ parametrizes the nonlinearity, stemming from $s$-wave interactions, and $\kbar$ denotes the effective Planck's constant.
 It arises naturally when rescaling all variables in the GPE to be dimensionless \cite{martin:matthew:chip}, and indicates how ``quantum'' the system is, with $\kbar\rightarrow0$ being the classical limit.
   
\begin{figure}
\centering
\epsfig{file={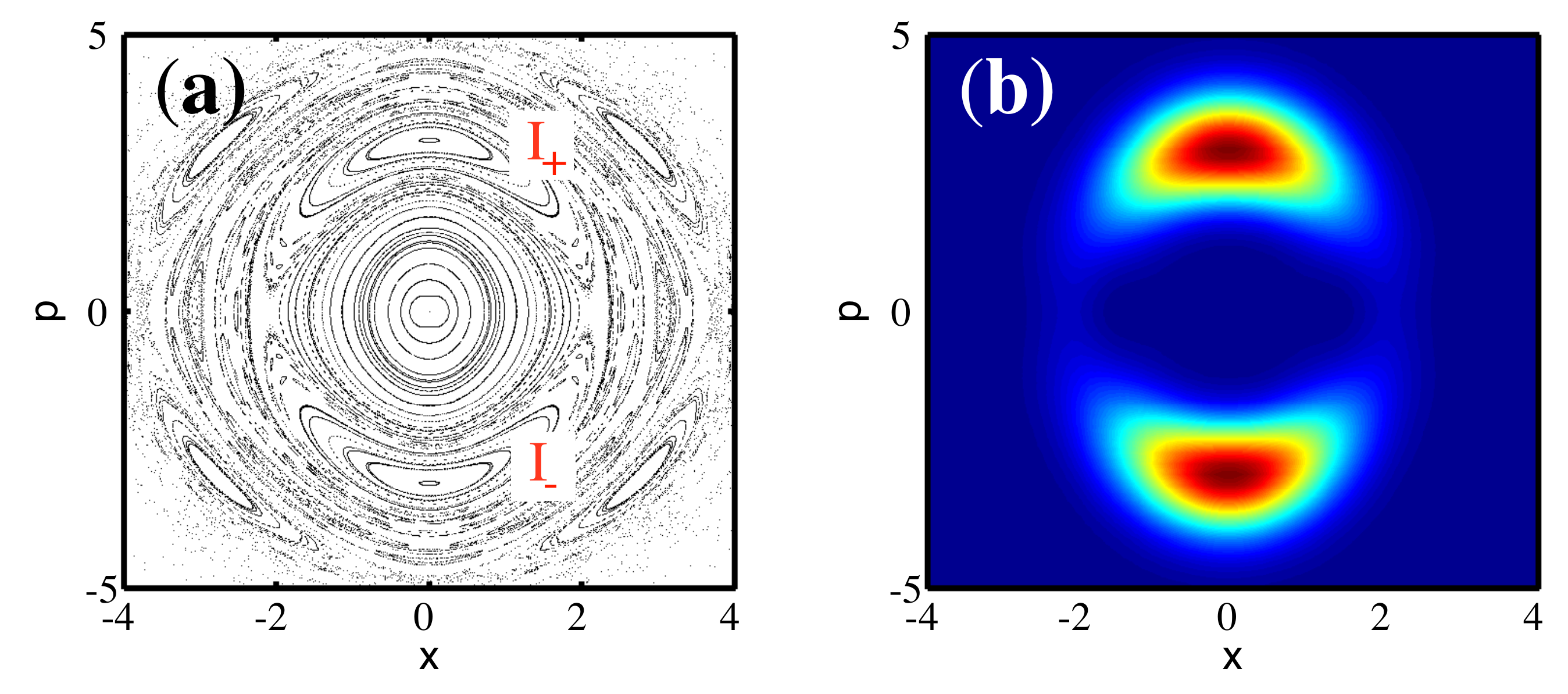},width=0.98\columnwidth} 
\caption{(color online) (a) Poincar\'e section for {the classical Hamiltonian  \eref{hamiltonian} with}  $\kappa=2.3$, $\epsilon=0.3$, showing regular islands $I_+$, $I_-$ separated by a region of chaos.  (b) Husimi function $Q(x,p)$ of the even tunnelling Floquet state: $Q(x,p)[\Psi]=\left|\langle \alpha | \Psi \rangle  \right|^{2}/{(2\pi\kbar)}$ for $\kbar=0.5$, where $|\alpha\rangle$ is a coherent state centered on momentum $p$ and position $x$.
\label{poincare_example}}
\end{figure}

The classical system \bref{hamiltonian} is integrable for $\epsilon=0$. The Kolmogorov-Arnol'd-Moser (KAM) theorem~\cite{book:reichl} states that regular regions of motion persist in phase-space for $\epsilon>0$, but become increasingly destroyed as $\epsilon$ is increased~\cite{martin:matthew:chip}. An example is shown in the Poincar\'e section of \frefp{poincare_example}{a}, where co-ordinates of classical motion from a large range of initial conditions are plotted stroboscopically, i.e.~at times $t=2\pi n$ for $n\in \mathbb{N}$. A key feature is the two large period-one islands of regular motion $I_\pm$, traced by trajectories of atoms moving in phase with the modulation of the potential~\cite{exp:hensinger}. The KAM theorem forbids classical {trajectories} connecting these islands.

Quantum mechanics, however, allows tunnelling to occur between {the period-one} islands. Consider the \emph{linear} Schr\"odinger equation ($U$=0) obtained from the quantized form of the Hamiltonian \bref{hamiltonian}. To relate the quantum dynamics of the modulated system to the classical phase-space, we use \emph{Floquet states}~\cite{book:reichl}, denoted $|u_{n}\rangle$, that are invariant up to a phase under time evolution through one modulation of period $T$, and can hence be found as eigenvectors of the time evolution operator: $\hat{U}(0,T)|u_{n}\rangle=\exp{[-i\lambda_{n}T/\kbar]}|u_{n}\rangle$~\cite{martin:matthew:chip}. The operator $\hat{U}(0,T)$ evolves the wave function from time $t=0$ to $t=T$. As $\hat{U}$ is unitary, the \emph{quasi-energy} $\lambda_{n}$ is real.
The period-one islands of regular motion occur in the Floquet spectrum as a pair of states that are even/odd respectively under the transformation $p\rightarrow-p$, and have support on both islands, as shown by the phase space Husimi function in \fref{poincare_example}(b).
 We will label these \emph{linear tunnelling states} $|u_{\textrm e}\rangle$ (even) and $|u_{\textrm o}\rangle$ (odd).

An atomic wavepacket that is initially localized on a single period-one island is a superposition of tunnelling states: 
{$|u_{\pm}(0)\rangle=[|u_{\textrm e}(0)\rangle \pm i |u_{\textrm o}(0)\rangle]/\sqrt{2}$} \cite{explain:i}, where $|u_{+}\rangle$ is located on the island with $p>0$. Using the time evolution of Floquet states, we have {$\hat{U}(0,nT)|u_{\pm}(0)\rangle=e^{-i \lambda_{\textrm e} n T/\kbar}  \left[ |u_{\textrm e}(0)\rangle+ i e^{i (\lambda_{\textrm e}-\lambda_{\textrm o}) n T/\kbar} |u_{\textrm o}(0)\rangle \right]/\sqrt{2}$}.
This gives rise to quantum tunnelling. Its experimental signature is a {classically forbidden} periodic reversal of the stroboscopically sampled atomic momentum as observed in~\cite{exp:hensinger,exp:raizen}. The quasi-energy splitting of the odd and even tunnelling states determines the \emph{linear} period of dynamical tunnelling: $\sub{T}{lin}=2\pi\kbar/|\lambda_{\textrm e}-\lambda_{\textrm o}|$.


For $U>0$ the problem is nonlinear, and  we cannot construct the operator $\hat{U}$ from the evolution of a set of basis states. Instead, we find nonlinear Floquet states~\cite{holthaus:NLfloq}  that are solutions $\phi_{n}$ of
\begin{align}
\left[ -\frac{\kbar^{2}}{2}\frac{\partial^{2}}{\partial x^{2}} + V(x,t) + U | \phi_{n} |^{2} -i\kbar \frac{\partial}{\partial t}\right] \phi_{n} =E_n\phi_{n}, 
\label{nonlin_Floqeqn}
\end{align}
periodic in the time dimension: $\phi_n(x,t)=\phi_n(x,t+2\pi)$ and vanishing for $x\rightarrow \pm \infty$. A state $\phi_{n}(x,0)$ will reform after one driving period of evolution with the GPE, up to a phase $-E_{n} T/\kbar$, analogous to the linear case \cite{footnote:conjgradsolution}. We  only consider the even (odd) nonlinear Floquet states localized on the islands, labelled $|\phi_{\textrm e}\rangle$ ($|\phi_{\textrm o}\rangle$).

\begin{figure}
\centering
\epsfig{file={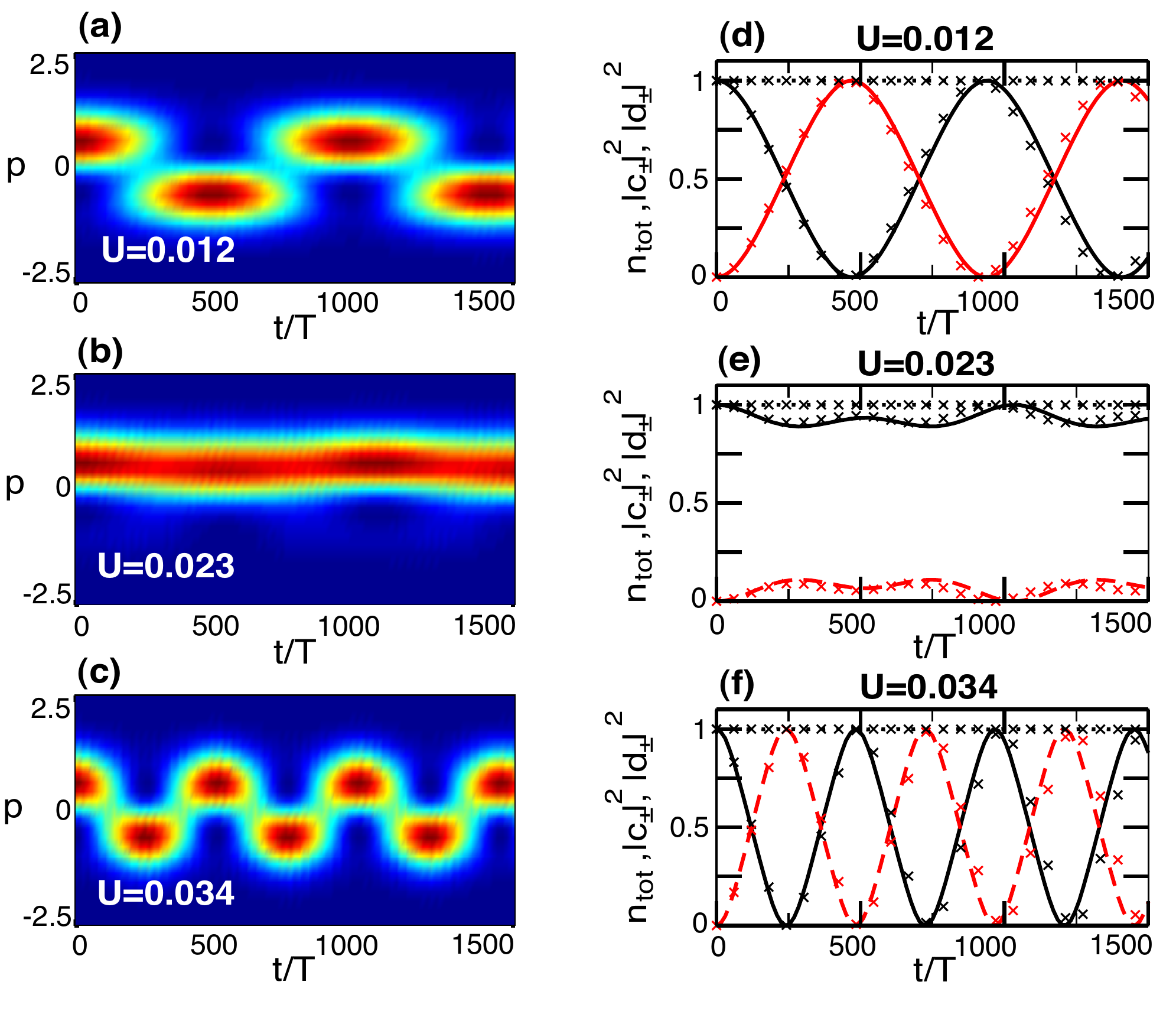},width=0.98\columnwidth} 
\caption{(color online) (a--c) Stroboscopic momentum space density showing the transition from (a) dynamical tunnelling (DT) to (b) MQST and (c) back to DT, with  
{$\kappa=1.3$, $\epsilon=0.2$} and {$U=\{1.2,2.3,3.4\}\times10^{-2}$} respectively. (d--f) Comparison of GPE Floquet state populations {$d_{\pm}(t)=\int dx \phi_{\pm}^{*}(x,t) \Psi(x,t)$} (lines) with two-mode model results [crosses, {$|c_{\pm}|^2$} {see \erefs{definition:coefficients}{twomodeqns}}]. Black-solid: $|d_{+}|^{2}$, red-dashed: $|d_{-}|^{2}$, black-dotted: {$\sub{n}{tot}\equiv$}$|d_{+}|^{2} + |d_{-}|^{2}$, black crosses $|c_{+}|^{2}$, red crosses: $|c_{-}|^{2}$. 
\label{dyntunnel_momspace}}
\end{figure}

Using $|\phi_{\textrm e/o}\rangle$ we simulate dynamical tunnelling with the GPE and $U>0$. We choose {$\kappa=1.3$, $\epsilon=0.2$} and $\kbar=0.5$, solving for nonlinear Floquet states up to {$U=2$}. 
We begin simulations in the state {$|\Psi\rangle= |\phi_{+}\rangle=[|\phi_{\textrm e}\rangle+i |\phi_{\textrm o}\rangle]/\sqrt{2}$} {\cite{explain:i}}, and evolve it with the GPE  for {$1500$} modulation periods.  We sample the momentum space wave function of the BEC  once every driving period $T$, {with} the results  shown in \fref{dyntunnel_momspace}.  

For $U=0.012$ we can see dynamical tunnelling despite the nonlinearity  {[\fref{dyntunnel_momspace}(a)]}. Its hallmark is a complete reversal of the system momentum on time-scales as long as about $500$ modulation periods. This tunnelling period is roughly three times longer than in the non-interacting case.

\begin{figure}
\centering
\epsfig{file={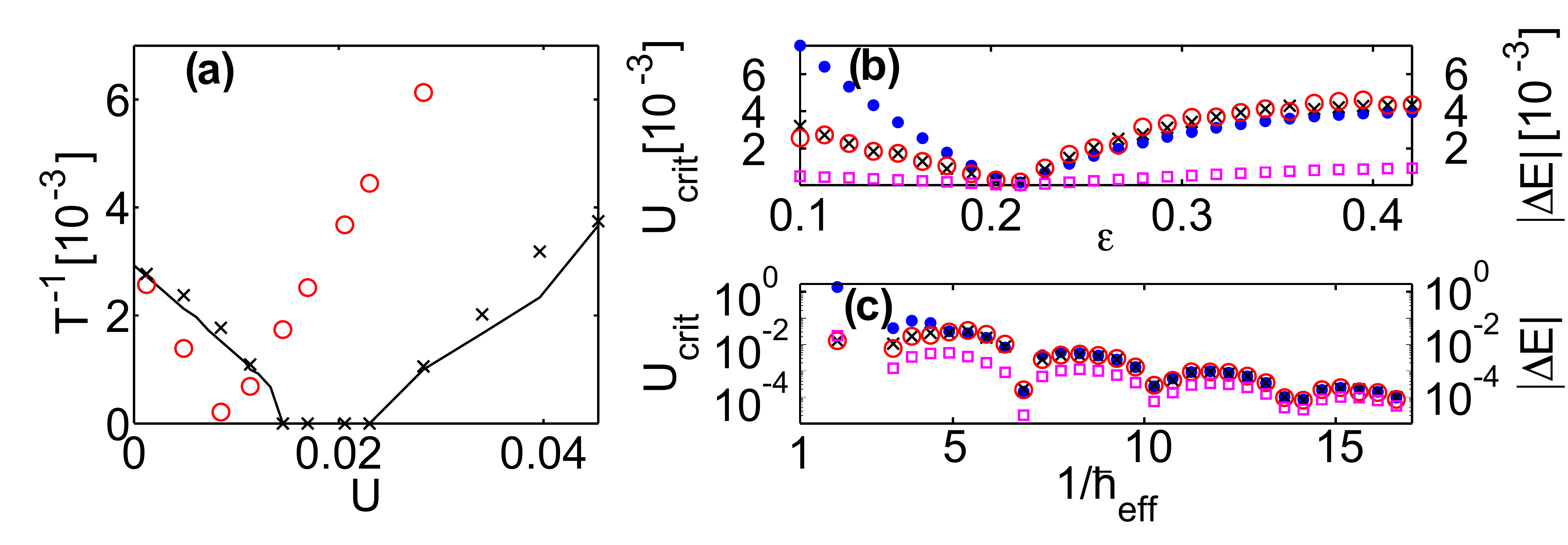},width=0.98\columnwidth} 
\caption{
(color online) (a) Dependence of the tunnelling rate $T^{-1}$ on the nonlinearity $U$, showing intermittent MQST for $U\in[1.4,2.2]\times10^{-2}$, with {$\kappa=1.3$, $\epsilon=0.2$} as in \fref{dyntunnel_momspace}. Black-solid: GPE solution, crosses: two-mode-model [\eref{twomodeqns}] with 
overlap coefficients from \eref{coupling:overlaps}. Red open circles: $\sub{T}{nl}^{-1}=|E_{\textrm e}(U) -E_{\textrm o}(U)|/2\pi\kbar$. 
{(b,c) Critical nonlinearity and tunnel splitting for $\kappa=1.2$ (b) as function of $\epsilon$ for $\kbar=0.15$, and (c) and as function of $\kbar$ for $\epsilon=0.2$. 
Blue dots: estimates from linear Floquet states using $\sub{U}{\textrm{crit}}=2 |\Delta E|/|\Lambda_{0}|$. 
Red open circles: nonlinear two-mode model, using \eref{condition}. Black crosses: extracted from full GPE simulations. Magenta squares: energy difference of Floquet states $|{\Delta}E|$.
}
\label{highUtunnel}}
\end{figure}

For $U=0.023$ complete momentum reversal no longer takes place, and the population becomes trapped in phase-space {[\fref{dyntunnel_momspace}(b)]}. This phenomenon is analogous to the cessation of inter-well tunnelling due to MQST in a bosonic Josephson junction~\cite{smerzi:mqst,oberthaler:exp1,holthaus:NLfloq,joel:MQST}. 
For $U=0.034$ we surprisingly find that tunnelling returns [\fref{dyntunnel_momspace}(c), see also \fref{highUtunnel}(a)]. For this parameter set it then persists for all nonlinearities that we modelled, as high as $U=2$ --- an effect that is not seen in the bosonic Josephson junction. In contrast, for many other parameter sets, tunnelling remains suppressed as $U$ is increased beyond the first onset of trapping. An extensive survey of parameter-space will be presented in Ref.~\cite{us:longpaper}.


To understand these results, we derive a two-mode model based on the nonlinear Floquet states. We assume that the time-dependent solution of the GPE can be approximated by two equivalent expressions 
\begin{subequations}
\begin{align}
\psi(x,t)&=c_{+}(t)\phi_{+}(x,t) + c_{-}(t)\phi_{-}(x,t),
\\
\psi(x,t)&=c_{\textrm e}(t)\phi_{\textrm e}(x,t) + c_{\textrm o}(t)\phi_{\textrm o}(x,t).
\label{definition:evenoddexpansion}
\end{align}
\label{definition:coefficients}
\end{subequations}
Members of both pairs are orthogonal by symmetry. 
We next insert \eref{definition:evenoddexpansion} into the GPE,  {make use of} \eref{nonlin_Floqeqn}, project out the equations of motion for $c_{\textrm e}(t)$ and $c_{\textrm o}(t)$, and finally change basis to $c_{+}(t)$ and $c_{-}(t)$. After defining {$\bar{E}=(E_{\textrm e} + E_{\textrm o})/2$, $\Delta E=E_{\textrm e} - E_{\textrm o}$ and coupling coefficients $U_{ij}$ with $\{i,j\} \in \{\textrm e,\textrm o\}$ and $A_{\textrm{eo}}$:}
\begin{subequations}
\begin{align}
U_{ij}(t)&=U\int dx |\phi_{i}(x,t)|^{2}|\phi_{j}(x,t)|^{2},
\\
A_{\textrm{eo}}(t)&=U\int dx \phi_{\textrm e}^{2}(x,t)\phi_{\textrm o}^{*2}(x,t),
\end{align}
\label{coupling:overlaps}
\end{subequations}
which are periodic in time with period $T$, we obtain:
\begin{eqnarray}
i\kbar \frac{\partial }{\partial t} c_{\pm}
&=&\bar{E}c_{\pm} + \Delta E c_{\mp}/2 
\CR
&+&\real{A_{\textrm{eo}}}|c_{\mp}|^{2}c_{\pm} - i \imag{A_{\textrm{eo}}} |c_{\pm}|^{2}c_{\mp}
\CR
&+&[U_{\textrm{eo}}-\real{A_{\textrm{eo}}}/2 -U_{\textrm{ee}}/4-U_{\textrm{oo}}/4 ]|c_{\pm}|^{2}c_{\pm}
\CR
&+&[i \imag{A_{\textrm{eo}}}/2 - U_{\textrm{ee}}/4 +U_{\textrm{oo}}/4]|c_{\mp}|^{2}c_{\mp}
\CR
&+&[U_{\textrm{ee}}/4 + U_{\textrm{oo}}/4 - U_{\textrm{eo}}- \real{A_{\textrm{eo}}}/2]c_{\mp}^{2}c_{\pm}^{*}
\CR
&+&[i \imag{A_{\textrm{eo}}}/2 + U_{\textrm{ee}}/4 - U_{\textrm{oo}}/4]c_{\pm}^{2}c_{\mp}^{*}.
\label{twomodeqns}
\end{eqnarray}
To test the model, we extract the populations of the modes $\phi_{\pm}(x,t)$ as a function of time  from the full simulations of the GPE, and compare them with the predictions of the two-mode model in Fig.~\ref{dyntunnel_momspace}(d--f).   In Fig.~\ref{highUtunnel}(a) we compare the tunnelling period of the full GPE against the two-mode model as a function of the nonlinearity $U$. The results demonstrate excellent agreement for these parameters.  

To analyse self-trapping, we consider the population imbalance $z=N_{+}-N_{-}$ and relative phase $\varphi=\theta_{-}-\theta_{+}$, where  $c_{\pm}=\sqrt{N_{\pm}}e^{i\theta_{\pm}}$ with $N_{\pm}, \theta_{\pm} \in \mathbb{R}$, following~\cite{smerzi:mqst}. 
For an analytical treatment, we replace the coefficients \bref{coupling:overlaps} by their average, e.g.~$U_{ij}(t)\rightarrow \bar{U}_{ij}=\frac{1}{T}\int_{0}^{T}U_{ij}(t)dt$, since tunnelling takes place on longer timescales.
It can be shown that $\imag{\bar{A}_{\textrm{eo}}}=0$ \cite{us:longpaper}.
The equations of motion for $z$ and $\varphi$, following from \eref{twomodeqns}, could be derived from the effective Hamiltonian
\begin{align}
\sub{H}{eff}&=\frac{\Lambda}{2} (1-\zeta) +\alpha \zeta^{\frac{1}{2}}\cos{(\varphi)} + \beta \zeta\cos{(2\varphi)},
\label{effhamiltonian}
\end{align}
where $\zeta=1-z^{2}$ and $\Lambda=(\bar{U}_{\textrm{ee}}+\bar{U}_{\textrm{oo}})/4 +3\real{\bar{A}_{\textrm{eo}}}/2-\bar{U}_{\textrm{eo}}$, $\alpha=(\bar{U}_{\textrm{ee}}-\bar{U}_{\textrm{oo}})/2 - \Delta E$, $\beta=\bar{U}_{\textrm{eo}}/2 -(\bar{U}_{\textrm{ee}}+\bar{U}_{\textrm{oo}})/8 + \real{\bar{A}_{\textrm{eo}} } /4$. For $\alpha=1$ and $\beta=0$ \eref{effhamiltonian} simplifies to the Hamiltonian of \rref{smerzi:mqst}, which analysed MQST for a BEC in a spatial double-well potential.

{Following~\cite{smerzi:mqst}} we can find Hamiltonian parameters for which dynamical tunnelling cannot occur. 
Starting from $z(0)=1$, energy conservation requires that for $z(t) = 0$ at some time $t$, there must exist a solution to
\begin{align}
\frac{1}{2}\Lambda&=\alpha \cos{[\varphi(t)]} + \beta \cos{[2\varphi(t)]}.
\label{condition}
\end{align}
The atoms are self-trapped when this equation cannot be fulfilled for any $\varphi(t)$. If we assume $|\Delta E| \gg |\beta|$ and $|\Delta E| \gg|(\bar{U}_{\textrm{ee}}-\bar{U}_{\textrm{oo}})/2|$, empirically justified in most cases, we find that tunnelling is impossible if $U>\sub{U}{crit}=2 |\Delta E|/|\Lambda_{0}|$. Here $\Lambda_{0}=\Lambda/U$ is an overlap integral between Floquet states that no longer explicitly depends on $U$, but {does so} implicitly through the shape of $|\phi_{i}(x,t)|^{2}$.
We can then estimate the critical nonlinearity for self-trapping from the \emph{linear} Floquet states, as they are generally very similar to the nonlinear Floquet states for $U<\sub{U}{\textrm{crit}}$. Instead of \eref{coupling:overlaps}, we then consider:
$
\tilde{U}_{ij}(t)=U\int dx |u_{i}(x,t)|^{2}|u_{j}(x,t)|^{2}$ and
$
\tilde{A}_{\textrm{eo}}(t)=U\int dx \:\: u_{\textrm e}^{2}(x,t)u_{\textrm o}^{*2}(x,t)$.
%
%
%

Equation~(\ref{condition}) does not always predict self-trapping. For  $\beta\ll\alpha$ (usually fulfilled) the self-trapping condition is
\begin{align}
1<\left| \frac{\Lambda}{2\alpha}  \right| =
\left| 
\frac{    (\bar{U}_{\textrm{ee}} + \bar{U}_{\textrm{oo}} )/4 -\bar{U}_{\textrm{oe}} + 3\real{  \bar{A}_{\textrm{eo}}  }/2    }
{\bar{U}_{\textrm{ee}}-\bar{U}_{\textrm{oo}} - \Delta E} 
\right|.
\label{condition2}
\end{align}
Aside from $\Delta E$ in the denominator, all terms in the fraction on the RHS are proportional to the nonlinearity $U$.  For $ U \gg \Delta E$ the nonlinearity then cancels out, and the condition \bref{condition2} depends only on the overlap integrals $\bar{U}_{ij}/U$ and $\bar{A}_{\textrm{eo}}/U$. These again are only weakly dependent on the nonlinearity $U$ through the \emph{shape} of $|\phi_{i}(x,t)|^{2}$. In particular for parameters where $\phi_{o}(x,t)$ and $\phi_{e}(x,t)$ have a significant difference in mean interaction energy, $|\bar{U}_{\textrm{ee}}-\bar{U}_{\textrm{oo}}|$, we will expect to see a reappearance of tunnelling at large $U$. This occurs for values of $\kappa \lesssim 2$; We find for $\kappa \gtrsim 2$ that $\bar{U}_{\textrm{ee}} \approx \bar{U}_{\textrm{oo}}$ \cite{us:longpaper}. An example without trapping at large $U$ is illustrated in \fref{highUtunnel}(a). The reappearance of tunnelling, a striking difference to the spatial double-well case, arises because the nonlinearity here affects both the self-energy of each tunnelling mode \emph{and} the effective mode coupling. 

In \fref{highUtunnel}(b--c) we plot the dependence of $\sub{U}{crit}$ on the driving amplitude $\epsilon$ and inverse effective Plank's constant $1/\kbar$, comparing $\sub{U}{crit}=2 |\Delta E|/|\Lambda_{0}|$ with a direct extraction from \eref{twomodeqns} and from the GPE. All models are in excellent agreement over a wide range of parameters. Plots of $\sub{U}{crit}$ directly reflect the groove structure also present in $|\Delta E|\sim 1/\sub{T}{lin}$ \cite{martin:matthew:chip}, indicating only minor changes in the coefficients $\Lambda_{0}$. A complete analysis of these parameter variations will be presented elsewhere~\cite{us:longpaper}.

The 1D nonlinearity $U$  can be related to experimental parameters by accounting for the details of the confinement geometry~\cite{martin:matthew:chip}. For example, for $\kappa=2.3$, $\epsilon=0.3$ we find $U_{\textrm{crit}}=0.004$ (onset of trapping) for $N=8$ atoms.  However, tunnelling will occur for $U=2$ with $\kappa=1.3$, corresponding to  $N=4590$ atoms \cite{footnote:numbers}. These disparate values for $N$ highlight the importance of our results for any experimental realisation of dynamical atom-chip tunnelling. We note that the large $\kbar$ used here would require {challengingly} tight trapping potentials~\cite{martin:matthew:chip,footnote:numbers}. {These in turn make experiments with large $U$ more realistic.}

In summary, we have demonstrated that the analogue of macroscopic quantum self-trapping in a bosonic Josephson junction exists in the dynamical tunnelling of BECs. However, we have discovered parameter regimes where MQST is lifted for large nonlinearities. We have shown 
that most of these features are reproduced by the dynamics of a simple two-mode model. An interesting extension of our work would be {to consider the quantum many-body} two-mode model, using methods of Refs.~\cite{weiss:meanfield_versus_MB,holthaus:coh_control:EPJB,heimsoth:orbitaljosephson}, or considering heating effects that can result from nonlinearities in the presence of driving~\cite{zhang:instability}.

\acknowledgments
We would like to thank P.~B.~Blakie, M.~Lenz and S.~Holt for assistance with the computer code.
This research was supported under the Australian Research Council's Discovery Projects funding scheme (DP0343094, DP0985142, DP1094025).

\bibliography{dyntunnel_short_new}

\begin{thebibliography}{41}
\bibitem[{exp()}]{explain:i}
\bibinfo{note}{{ The phase-factor $i$ results from our choice of phase of the
  potential modulation, as $|u_+(t=0)\rangle$ describes a particle with $v>0$
  and hence non-vanishing phase gradient }}.
\bibitem[{foo({\natexlab{a}})}]{footnote:conjgradsolution}
\bibinfo{note}{{We solve \eref{nonlin_Floqeqn} using conjugate gradient
  techniques, treating time as a second ``spatial'' dimension}}.
\bibitem[{foo({\natexlab{b}})}]{footnote:numbers}
\bibinfo{note}{{For the reduction of dimensions from 3D to 1D described in
  \cite{martin:matthew:chip}, the total 3D atom number is $N=L_z
  U/(2\sqrt{2\pi} a_s \kbar^{3/2}\kappa^{1/4})\sqrt{\omega_x/\omega_y}$, where
  $a_{s}$ is the three-dimensional scattering length, $\omega_x$ the strength
  of the chip trap, $\omega_y$ the potential strength in the frozen
  $y$-dimension and $L_z$ the condensate size in the long, uniform
  $z$-direction (see \cite{martin:matthew:chip}). For the $N$ quoted, we assume
  $L_z=50$ $\mu$m, $\omega_x=(2\pi)\times10$ kHz, $\omega_y=5\omega_x$ and
  $a_s=5.5$ nm ($^{87}$Rb).}}
\end{thebibliography}
\end{document}